\documentclass[aps,preprint,prl,showpacs,superscriptaddress]{revtex4}

\usepackage{amsmath}
\usepackage{amssymb}
\usepackage{bm}
\usepackage{graphicx}
\usepackage{dcolumn}

\begin{document}


\title{Formation and Evolution of Single Molecule Junctions}

\author{M. Kamenetska}
\affiliation{Department of Applied Physics and Applied Mathematics, Columbia University, New York, NY}
\affiliation{Center for Electron Transport in Molecular Nanostructures, Columbia University, New York, NY}
\author{M. Koentopp}
\affiliation{Center for Electron Transport in Molecular Nanostructures, Columbia University, New York, NY}
\author{A. C. Whalley}
\affiliation{Department of Chemistry, Columbia University, New York, NY}
\author{Y. S. Park}
\affiliation{Department of Chemistry, Columbia University, New York, NY}
\author{M. L. Steigerwald}
\affiliation{Department of Chemistry, Columbia University, New York, NY}
\affiliation{Center for Electron Transport in Molecular Nanostructures, Columbia University, New York, NY}
\author{C. Nuckolls}
\affiliation{Department of Chemistry, Columbia University, New York, NY}
\affiliation{Center for Electron Transport in Molecular Nanostructures, Columbia University, New York, NY}
\author{M. S. Hybertsen}
\email{mhyberts@bnl.gov}
\affiliation{Center for Functional Nanomaterials, Brookhaven National Laboratory, Upton, NY}
\author{L. Venkataraman}
\email{lv2117@columbia.edu}
\affiliation{Department of Applied Physics and Applied Mathematics, Columbia University, New York, NY}
\affiliation{Center for Electron Transport in Molecular Nanostructures, Columbia University, New York, NY}


\begin{abstract}
We analyze the formation and evolution statistics of single molecule junctions bonded to gold electrodes using amine, methyl sulfide and dimethyl phosphine link groups by measuring conductance as a function of junction elongation. For each link, maximum elongation and formation probability increase with molecular length, strongly suggesting that processes other than just metal-molecule bond breakage play a key role in junction evolution under stress. Density functional theory calculations of adiabatic trajectories show sequences of atomic-scale changes in junction structure, including shifts in attachment point, that account for the long conductance plateau lengths observed.
\end{abstract}

\date{\today}


\maketitle


Over the past decade, the field of molecular scale electronics has come a long way towards elucidating and characterizing intrinsic molecular properties that affect transport. The electronic properties of single molecules attached to metal electrodes have been measured successfully by elongating and breaking nanometer scale wires in an environment of molecules using mechanically controlled break junctions and scanning tunneling microscopes (STM) \cite{ref1,ref2,ref3,ref4}. Typically, the focus of these measurements has been on conductance and current-voltage characteristics. However, because the physical structure of a single, nanoscale junction, such as an Au point-contact, has only rarely been directly observed \cite{ref5}, fundamental questions regarding link bond formation and junction evolution under stress remain to be answered.

Here, we analyze a statistically significant sample of single molecule junctions made using a simplified STM, where the junction conductance is recorded as a function of the relative tip-sample displacement. These conductance versus displacement traces, measured with amine (NH$_2$), dimethyl phosphine (PMe$_2$), and methyl sulfide (SMe) links that bind selectively to gold \cite{ref6}, show plateaus that appear during elongation, providing a signature of junction formation. The length of these plateaus for different molecules probes the amount of elongation a junction can sustain without breaking apart. Our results show that, across all end groups, longer molecules form longer conductance plateaus and have a higher probability of forming a junction, while changes in applied bias voltage or elongation speed have no discernable effect \cite{ref7}. Density functional theory (DFT) based ab-initio calculations simulating the junction elongation process for the NH$_2$ and PMe$_2$ links show clearly that the long steps result from multiple processes including changes in the molecular binding site, changes in the gold electrode structure, molecular rearrangements, as well as bond-breakage. Bond-breakage contributes only a small fraction of the total junction elongation length. Furthermore, the zero-bias transmission does not change significantly upon changes in molecular binding site or gold electrode structure, consistent with the narrow conductance peaks seen in the measured conductance histograms.

We form single molecule junctions by breaking Au point contacts with a simplified STM in a 1,2,4-trichlorobenzene solution of the molecules, as detailed previously \cite{ref8}. The inset of Figure 1a compares individual conductance traces for 1,4-butanediamine (\textbf{1}), 1,6-hexanediamine (\textbf{2}) and 1,4-bis (dimethylphosphino)butane (\textbf{3}), measured under bias of 25 mV with a pulling speed of 15 nm/s. These sample traces show plateaus with a molecule dependent conductance and length; junctions of butane with the PMe$_2$ links (\textbf{3}) have the longest plateaus, sustaining the largest elongation while those of butane with NH$_2$ links (\textbf{1}) have the smallest. Figure 1a shows the corresponding normalized conductance histograms generated, without any data selection, by binning the conductance values from around 40,000 measured traces. The peak seen in these histograms reflects well the fact that molecular plateaus occur repeatedly at well-defined conductance values.

\begin{figure*}[t]
\centering

\includegraphics[bb=30 50 560 200,clip,scale=0.85]{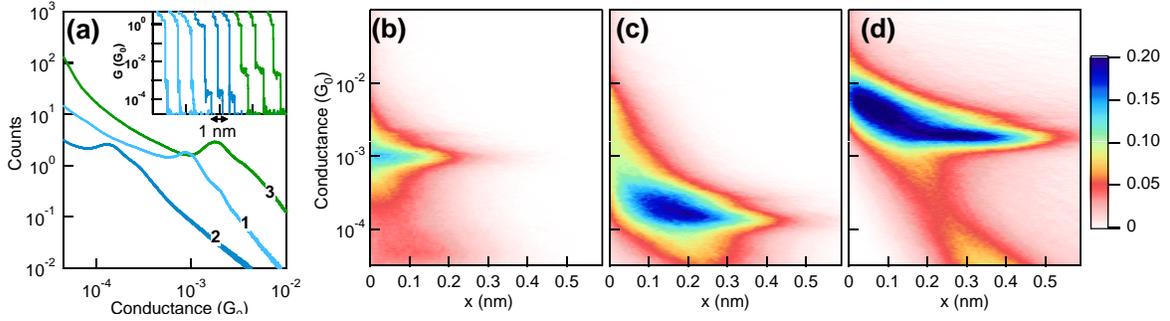}
\caption{
(a) Normalized linear conductance histograms of 1,4-butanediamine ({\bf 1}, bin size $10^{-5} G_0$), 1,6-hexanediamine ({\bf 2}, bin size $10^{-6} G_0$) and 1,4-bis(dimethylphosphino)butane ({\bf 3}, bin size $10^{-5}G_0$). Inset: Sample conductance traces (offset horizontally for clarity) that show a molecular step for each molecule. (b-d) Normalized 2D histograms for molecules {\bf 1}, {\bf 2} and {\bf 3} respectively showing counts as a function of conductance (y-axis) and displacement relative to G$_0$ gold-gold contact break (x-axis).
}
\end{figure*}

To distinguish differences in molecular plateau lengths, a two-dimensional (2D) histogram, retaining displacement information, is required \cite{ref9}. Since conductance plateaus occur in random locations along the displacement axis, we first set the origin of our displacement axis at the point in the conductance traces where the gold-gold contact breaks and the conductance drops below G$_0$. With this origin, (determined individually for each trace using an automated algorithm\cite{ref10}), 2D histograms are computed from all conductance traces, using linear bins along the positive displacement axis (x-axis) and log bins along the conductance axis (y-axis) for image clarity. The normalized 2D histograms for these three molecules show that the molecular conductance peak extends to approximately 2.5 \AA~along the x-axis for \textbf{1} (Figure 1b), 4.5 \AA~for \textbf{2} (Figure 1c), and 5.5 \AA~for \textbf{3} (Figure 1d). The comparison of  \textbf{1} and \textbf{2} provides statistical evidence that plateaus are indeed longer and more frequent for the longer molecule. In addition, we see that molecules with the PMe$_2$ (\textbf{3})link have longer plateaus when compared with the NH$_2$ link (\textbf{1}).

To quantify these trends, we generate histograms of plateau lengths, for a series of alkanes with NH$_2$, SMe and PMe$_2$ links, by determining the length of the molecular conductance plateau for each measured trace with an automated algorithm \cite{ref10}. From such a histogram (Figure 2a), we determine the ``longest'' plateau for each molecule measured, defined as the 95th percentile of the distribution \cite{ref11}. In Figure 2b, we plot the ``longest'' step length as a function of the number of methylene (CH$_2$) groups on the alkane back-bone for all molecules measured. Comparing molecules with the same link group, we see a striking increase in junction elongation distance with molecule length. Linear fits to these data show a similar slope, but they are offset vertically, with the PMe$_2$ link having the largest intercept. This is consistent with our previous assertion that the elongation distance between the energy minimum configuration and the force maximum of a molecular junction increases from NH$_2$ to SMe to PMe$_2$ \cite{ref6}.

\begin{figure}[t]
\centering
\includegraphics[bb=30 50 600 280,clip,scale=0.45]{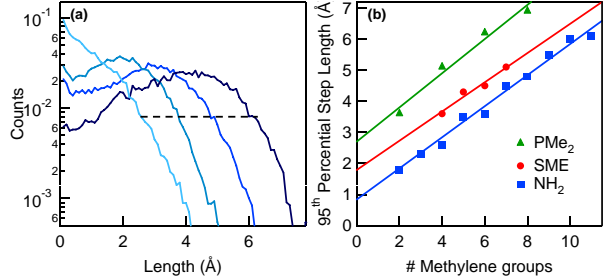}
\caption{
(a) Conductance step length distribution for diamino-alkanes with 4, 6, 8 and 11 methylene groups. Dashed line indicates the 95$^{th}$ percentile length for each distribution. (b) Conductance step length (95$^{th}$ percentile) for alkanes terminated with NH$_2$, SMe or PMe$_2$ link groups,
as a function the number of methylene (CH$_2$) groups in the chain.
}
\end{figure}

The fact that the plateau lengths depend on the molecule length suggests strongly that junctions can form with one (or both) of the link groups bonded to an Au atom away from the apex of the tip or substrate. Upon elongation, the binding site could move from one atom to the next, or the gold electrodes could deform under the pulling force. Such scenarios would imply that longer molecules have access to a larger number of binding sites on the Au tip and substrate. Indeed, we find that the fraction of traces with steps does increase systematically with molecule length from about 25\% for ethanediamine, to 65\% for butanediamine, 85\% for hexanediamine and about 95\% for 1,8-octanediamine.

\begin{figure}[t]
\centering
\includegraphics[bb=30 55 400 576,clip,scale=0.45]{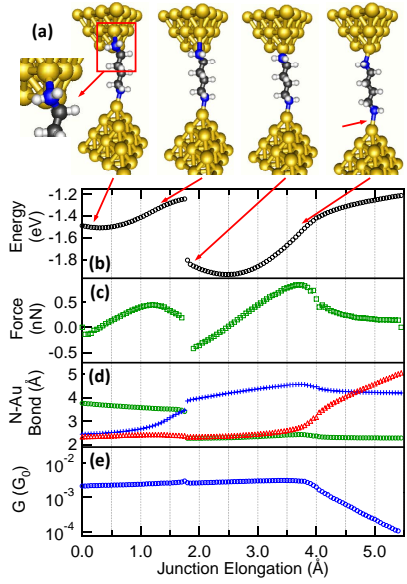}
\caption{
Calculated adiabatic 1,4-butanediamine model junction elongation trace.
(a) Relaxed junction geometries at four different positions during the trajectory.
(b) Binding energy relative to relaxed, isolated Au pyramids and the molecule.
(c) External applied force calculated
from the derivative of the junction energy with respect to elongation.
(d) N-Au bond lengths: upper N to the Au atom on the second layer (+),
to the Au atom on the bottom layer ($\square$),
and lower N-Au bond ($\triangle$).
(e) Junction conductance.
}
\end{figure}

To probe these different hypotheses, ab-initio calculations of adiabatic pulling traces were conducted for \textbf{1} and \textbf{3} with different initial link geometries. The contacts are modeled with Au pyramids (20 atoms each) with (111) surfaces. The tip atom on the top pyramid was moved to an adatom site on one facet.  We considered starting geometries in which one link group was bound to an atom on the edge of the top pyramid (Figure 3) or to the adatom on one of the faces (Figure 4). The back layer of Au atoms for the top and the back two layers of the bottom pyramid were held fixed with a bulk lattice parameter 4.08 \AA~(except where noted below). All other degrees of freedom were relaxed until all forces were less than 0.005~eV/\AA. The junction was then elongated in 0.05 \AA~steps by a shift of the bottom pyramid along the z-direction followed by full geometry optimization.

Total energy calculations and geometry optimization were performed with the quantum chemistry package TURBOMOLE v5.10 \cite{ref12,ref13,ref14,ref15,ref16}. A DFT approach was used with a generalized gradient approximation functional (PBE) \cite{ref17} and an optimized split valence basis set with polarization functions (designated def2-SVP) \cite{ref18,ref19,ref20}. The ballistic electron transmission through the junction was calculated with a Green's function approach applied to the composite electrode-molecule system and a simplified embedding self energy \cite{ref21,ref22,ref23}. The zero bias conductance is given by the transmission at the Fermi energy. The Green's function was based on the eigenstates from the DFT calculation. Test conductance calculations for 1,4-benzenediamine agreed with earlier results \cite{ref10}. While the DFT-derived frontier energy alignment results in systematic errors in the calculated conductance \cite{ref10,ref23a,ref24}, the errors are modest for alkanes because the electrode Fermi energy is roughly in the middle of the HOMO-LUMO gap \cite{ref25,ref26}.

We studied eight distinct butane junction structures, four with each link. Figure 3 shows an illustrative scenario where the NH$_2$ initially binds to the edge atom of the second layer of the upper pyramid \cite{ref27}. The NH$_2$ remains coordinated to this Au atom for about 1.5~\AA, with an increasing N-Au bond length. When it approaches a bridging geometry, the NH$_2$ abruptly jumps (at z=1.8\AA) to bind to the lower, corner Au atom on the tip. Following the jump, the N-Au bond lengths remain relatively constant up to about z=3.5~\AA, with the geometry adjusting through bond angle changes. Then the bottom N-Au starts to elongate and the maximum sustained force of 0.8~nN is observed near z=3.8~\AA. Up to z=3.8~\AA, the calculated conductance is consistent with a single step as it would be observed in the experiments. After z=3.8~\AA, the conductance decreases exponentially with N-Au bond elongation. The abrupt termination of experimental traces is consistent with energy cost to break the lower N-Au bond ($<$0.4 eV) at z=3.8 \AA~and thermal fluctuations on the millisecond time scale. Calculations starting from a similar junction with PMe$_2$ links showed more extensive Au electrode deformations, including plastic deformation of the tip region and extraction of short Au chains. The P-Au bond to the lower pyramid broke after a 5-7~\AA~elongation with a maximum sustained force around 1.4~nN.

\begin{figure}[t]
\centering
\includegraphics[bb=30 55 400 576,clip,scale=0.45]{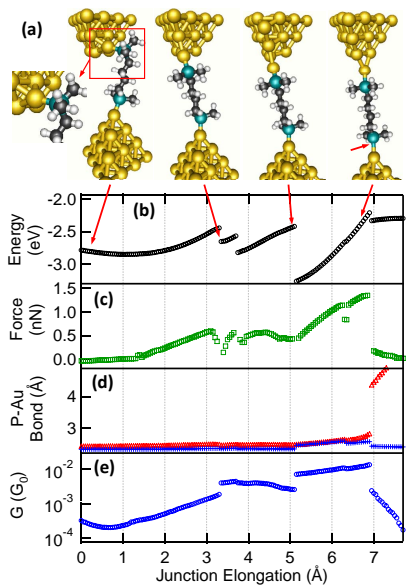}
\caption{
Calculated adiabatic 1,4-bis(dimethylphosphino)-butane model junction elongation trace.
(a) Relaxed junction geometries at four different positions during the trajectory.
(b) Binding energy.
(c) External applied force.
(d) P-Au bond length: upper (+), lower and bottom ($\triangle$) pyramids.
(e) Junction conductance.
}
\end{figure}

Figure 4 shows a different scenario with a PMe$_2$ link initially bound to the Au adatom tip face \cite{ref27}. There is an initial twist in the molecule ($\sim$0.2~eV energy cost) which could be realized in experiment due to constraints in available binding sites. The conductance thus starts low (2~-~3$\times10^{-4}$~G$_0$) as the electronic gateway state of the P-Au link is not aligned with the sigma states of the alkane backbone. Under stress, the molecule untwists, initially slowly and then with a rapid readjustment at around z=3.3~\AA, and conductance rises to $\sim 4\times10^{-3}$~G$_0$. At the same time, the Au adatom is dragged towards the edge of the pyramid, adopting a two-fold coordination against the pyramid edge. At z=3.7~\AA, the nearest Au corner atom in the back layer is freed. This stabilizes the upper Au pyramid against significant plastic distortion. From z=3.9 to 5.1~\AA, the Au atoms around the adatom distort and at z=5.1~\AA, the adatom abruptly jumps to the apex position with the conductance increasing to about $7\times10^{-3}$~G$_0$. The P-Au bonds and Au apex structures stretch modestly with the lower P-Au bond taking up most of the elongation until it breaks at z=6.9~\AA. The conductance then decreases exponentially. The maximum sustained force is again about 1.4~nN, similar to the measured Au-Au breaking force \cite{ref28}. Experimentally, such a trace would have an initial gap (low conductance value). Calculations done for similar junction structures with NH$_2$ links showed that the N-Au bond was only strong enough to pull the Au adatom up to a bridging position on the pyramid edge before the lower N-Au bond broke at a maximum sustained force of 0.8~nN.

An overview of all eight calculated adiabatic trajectories shows that the junction formation energy at local minima spans the range 1.1 - 1.6~eV per PMe$_2$-Au bond (15 minima) and 0.7 -  1.0~eV per NH$_2$-Au bond (8 minima). With the exception of regions with a twist in the molecule, the calculated conductance values undergo modest changes when the link attachment point shifts or the Au atoms near the link rearrange. The calculated conductance values for the NH$_2$ linked butane range from $1 - 3\times10^{-3}$ G$_0$ while those for the PMe$_2$ linked butane span a broader range ($1-12 \times 10^{-3}$ G$_0$). These are slightly larger than the experimental peak positions in Figure 1a at approximately $1\times10^{-3}$ G$_0$ (NH$_2$) and $2\times10^{-3}$ G$_0$ (PMe$_2$), though the ranges are consistent with the measured histogram widths.

Several fundamental points emerge from our experiments and calculations.  First, junctions can form with the link group bonded to an undercoordinated Au atom higher up on the electrode with similar binding energy and conductance.  This naturally explains the higher probability of junction formation observed for longer molecules.  Second, under stress, such junctions can evolve through different types of physical motion of the link, either by hopping of the link group from one available undercoordinated Au site to another (but never a bridge or hollow site), or by dragging an undercoordinated Au atom, thereby distorting the Au structure. During the course of this motion, the conductance can be relatively stable, consistent with a single step in the measured traces.  Third, the stronger interaction of the PMe$_2$ link group with the Au, compared with the NH$_2$ link, can extend the physical junction over a larger range of elongation, consistent with measurements. Finally, our results highlight the diversity of physical configurations that are probed when single molecule junctions are formed. The consistent measured conductance signatures that form the well defined steps derive from the chemical specificity of the donor-acceptor link motifs. The lone pair on the link atom (N from NH$_2$, S from SMe and P from PMe$_2$) coordinates a single Au atom on the electrode, thus local variations in electrode atomic structure and link geometry have only a modest influence on the electronic coupling.

We thank Ferdinand Evers for use of the advanced NEGF code. This work was supported in part by the Nanoscale Science and Engineering Initiative of the NSF (award numbers CHE-0117752 and CHE-0641532), the New York State Office of Science, Technology, and Academic Research (NYSTAR) and the NSF Career Award (CHE-07-44185). Part of this research was performed at the Center for Functional Nanomaterials, Brookhaven National Laboratory, supported by the U.S. Department of Energy, Office of Basic Energy Sciences, under contract number DE-AC02-98CH10886.


\begin{references}
\bibitem{ref11}
Although the 95th percentile is chosen arbitrarily,
it gives a good measure of a statistically significant
``longest'' step,
the geometrically relevant quantity probed here.
\bibitem{ref23}
With the 20 Au atom clusters, the self energy ($\Gamma = 2.7$ eV) is applied
to the back two planes.
The calculated conductance is approximately constant for $\Gamma = 1 - 5$ eV.
\end{references}
\end{document}